\documentclass[aps,prl,twocolumn,showpacs,groupedaddress,amsmath,amssymb]{revtex4}
\usepackage{graphicx}% Include figure files
\usepackage{color}
\usepackage{bm}

\begin{document}

\title{The Effect of non-Hermiticity on Adiabatic Elimination}
\author{Rahman Sharaf$^1$, Mojgan Dehghani$^2$, Sara Darbari$^1$, Hamidreza Ramezani$^2$} 
\email {hamidreza.ramezani@utrgv.edu}
\affiliation{$^1$Faculty of Electrical and Computer Engineering,Tarbiat Modares University,Tehran 14115, Iran\\
$^2$Department of Physics, The University of Texas Rio Grande Valley, Brownsville, TX 78520, USA}

\begin{abstract}
We investigate the influence of non-Hermiticity on the adiabatic elimination in coupled waveguides. We show that adiabatic elimination is not affected when the system is in parity-time symmetric phase. However, in the broken phase the eliminated waveguide loses its darkness namely its amplitude starts increasing, which means adiabatic elimination does not hold  in the broken phase. Our results can advance the control of the dynamics in coupled laser cavities, and help the design of controllable absorbers.

\end{abstract}


\maketitle
{\it Introduction--}The demand on designing miniaturized and on chip optical devices with new functionalities has been increased in the last few years. Many linear and nonlinear structures such as nano-size fabricated waveguides \cite{1}, spherical and disk micro-resonators \cite{2,disk}, photonic crystals \cite{eli}, and metamaterials \cite{Zheludev} has been proposed. On a related matter of finding new structures, non-Hermitian systems have recently become the center of attention in photonics \cite{ramy, nat10, flat}, electronics \cite{elec}, and acoustics \cite{prx,natcom,alu, newj}, due to their fascinating features and applications \cite{laser1,laser2,absorber}. Among the non-Hermitian systems, Parity and Time reversal (PT) symmetric systems are specifically important because of showing phase transition from the exact phase with real spectrum to the broken phase with complex spectrum \cite{bender}. The transition point is known as exceptional point or PT symmetry breaking point. At the exceptional point two or more eigenvectors of the system in a pairwise manner coalesce and their associated eigenvalues become degenerate. In each phase, many interesting features have been discovered and experimentally demonstrated. Among them are unidirectional invisibility \cite{zin, dim}, parity anomaly \cite{henning}, loss induced lasing \cite{yang}, protected bound state\cite{hen,alex}, and non-Hermiticity induced flat band \cite{flat}.

Despite all these achievements, designing the nano-size PT structures is very difficult. Practically, PT structures are composed of some gain and loss mechanisms that are judiciously distributed in the structure. Usually in such systems, there are some modes that are coupled with each other through coupling constants generated by the overlap integral of the field. In the absence of the couplings, the modes are degenerate and the degeneracy breaks by introducing the coupling. The level spacing between coupled modes is proportional to the coupling which dictates the bandwidth of the system. The introduction of the gain and loss reduces the level spacing until the system reaches the spontaneous symmetry breaking point where the level spacing becomes zero.  Thus, when the coupling between the modes is stronger, the bandwidth of the system becomes wider and consequently the exact phase of the PT system becomes larger. On the other hand, in nano-scale and densely packed subwavelength structures such as coupled single mode waveguides, where each mode exists in one waveguide, the fabricated waveguides are very close to each other \cite{xiangadi}. The PT version of the densely packed subwavelength waveguides involves gain and loss waveguides. While the loss is usually generated by metal coating \cite{salamo}, the gain is created using optical or electrical pumping. However, in subwavelength structures, the diffraction limit (in the case of optical pumping) or tunneling of electrons and diffusion current (in the case of electrical pumping) put a major barrier to advance the field of non-Hermitian photonics in subwavelength regime and to fabricate PT symmetric densely packed structures. Similar issues encounter when we deal with densely coupled laser cavities with different gain thresholds.
\begin{figure}
	\includegraphics[width=1\linewidth, angle=0]{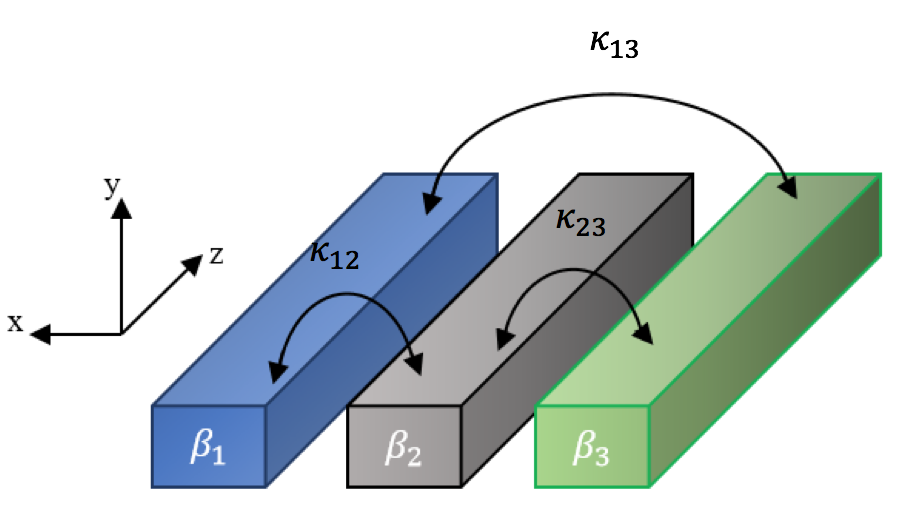}
	\caption{(Color online) Schematic of the trimer coupled waveguides. The green waveguide is a loss waveguide, the blue one is gain while the middle waveguide is considered to have no gain or loss. In AE, one wants to make the electric field intensity in the middle waveguide to be zero. \label{fig1}}
\end{figure}

Here we propose to bypass the above problem using the concept of adiabatic elimination (AE). Specifically, we derive the condition for AE in three coupled waveguides, one with gain, a centric waveguide with no gain or loss, and a loss waveguide. We derive the condition in which the system supports AE and can be effectively described by a $2\times2$ effective Hamiltonian. Such Hamiltonian leads us to identify the exact phase. Furthermore, we show that in the broken phase AE breaks. First, we analyze the beam dynamics in such a PT system and then show that while the passive waveguide in the exact phase can be eliminated, in the broken phase the field intensity exponentially increases in all the waveguides. Although we consider coupled waveguides, our approach can be generalized to other systems such as coupled electronic oscillators, coupled laser cavities, and absorbers.

{\it Model--} Let us consider three coupled waveguides as shown in Fig. (\ref{fig1}). The left (right) waveguide with propagation constant $\beta_{1(3)}=\beta_{1(3)}^{\prime}-(+)i\gamma$ is the gain (loss) waveguide, the middle one is a passive waveguide with $\beta_2=\beta_{2}^{\prime}$, where $\gamma$ identifies the value of the gain and loss. The couplings between the waveguides are given by $\kappa_{12}$, $\kappa_{23}$, and $\kappa_{13}$ as indicated in the Fig.(\ref{fig1}). The PT symmetry condition forces the following relations:
\begin{equation}
\beta_{1}^{\prime}=\beta_{3}^{\prime}\equiv \beta^{\prime} \quad \kappa_{12}=\kappa_{23}\equiv \kappa
\label{eq1}
\end{equation}
With a good approximation, we can write the couple mode equations describing the electric field propagation $\vec{E}=(\begin{array}{ccc}
E_1&E_2&E_3
\end{array})^T$in the system:
\begin{equation}
i\frac{d}{dz}\vec{E}=\left(\begin{array}{ccc}
\beta_1&\kappa&\kappa_{13}\\
\kappa^\ast&\beta_2&\kappa\\
\kappa_{13}^\ast&\kappa^\ast&\beta_3
\end{array}\right)\vec{E}.
\label{eqe}
\end{equation}
Writing the above equations in terms of electric field amplitudes $E_j=A_j \exp(i\beta_j z)$, we get
\begin{equation}
 i\frac{d}{dz}\vec{\psi}=H\vec{\psi},\quad \vec{\psi}=(\begin{array}{ccc}
 A_1&A_2&A_3
 \end{array})^T
 \label{eqd}
\end{equation}
where $z$ is the propagation direction, $A_{j=1,2,3}$ represents the electric field amplitude in gain ($j=1$), passive ($j=2$), and loss ($j=3$)waveguides, respectively. Moreover,
\begin{equation}
H=-\left(\begin{array}{ccc}
0 & \kappa_{12}e^{-i\Delta\beta_{12} z}&\kappa_{13}e^{-i\Delta\beta_{13} z}\\\kappa_{12}^\ast e^{i\Delta\beta_{12} z}&0&\kappa_{23}e^{-i\Delta\beta_{23} z}\\\kappa_{13}^\ast e^{i\Delta\beta_{13} z}&\kappa_{23}^\ast e^{i\Delta\beta_{23} z}&0
\end{array}\right)
\label{eq2}
\end{equation} 
where $\Delta\beta_{ij}=\beta_i-\beta_j$ and $\ast$ means conjugation.

\begin{figure}
	\includegraphics[width=1\linewidth, angle=0]{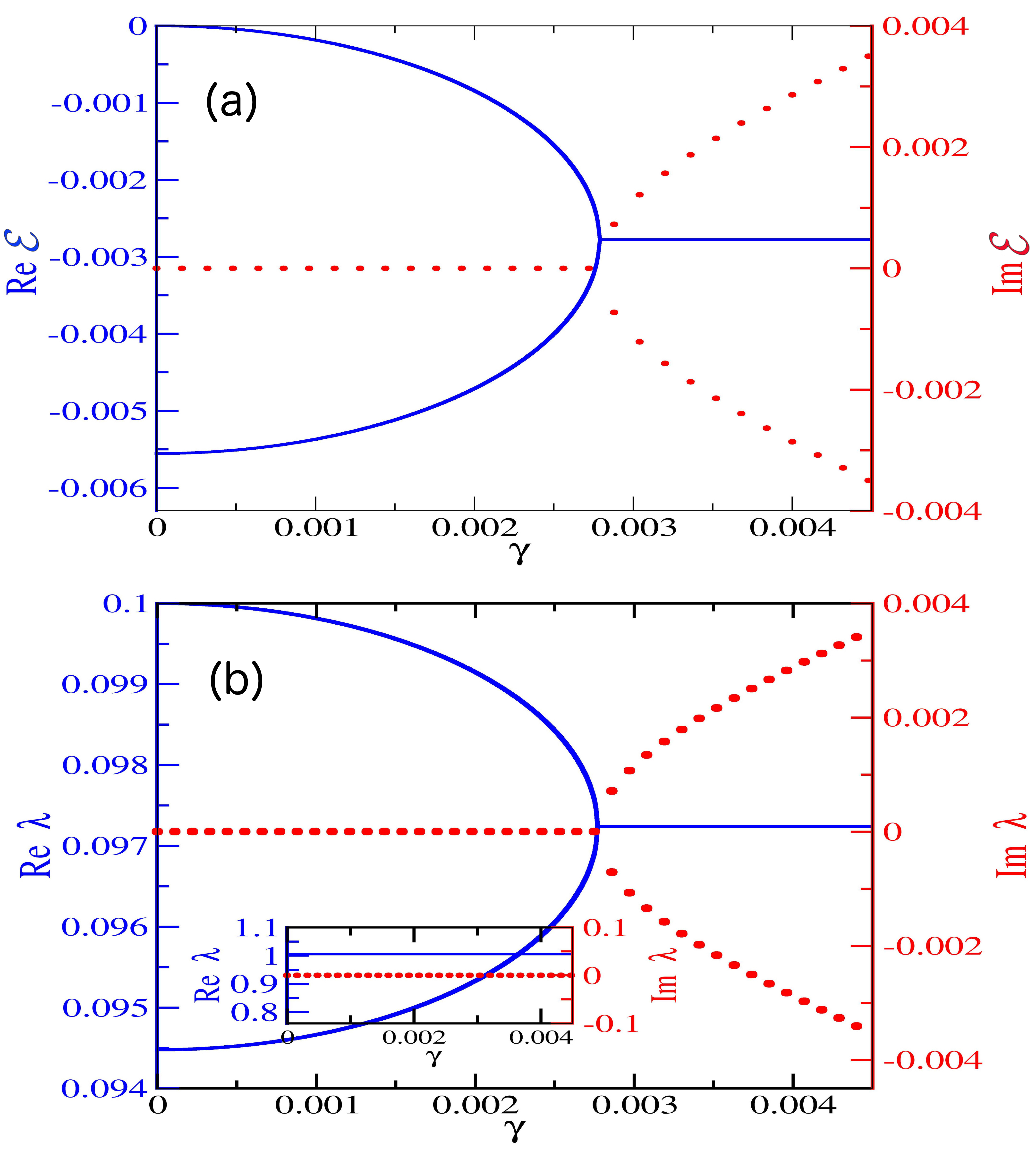}
	\caption{(Color online) (a) Real parts (blue curve) and imaginary parts (red dots) of the Eigenvalues of the $\mathcal{H}_{eff}$, vs. $\gamma$. (b) Spectrum of Eq.(\ref{eqe}), vs. $\gamma$. A comparison shows that the effective system after AE approximation accurately describes the exact and broken phase of the system.\label{fig2}}
\end{figure}
In the AE regime, we expect that the middle waveguide remains a dark state at any $z$. Thus, its intensity is not influenced by the field  of the gain and loss waveguides and remains constant. Consequently, we can integrate the middle row of the \ref{eq2}, which results in 
\begin{equation}
A_2=\frac{\kappa_{12}^\ast}{\Delta\beta_{12}}e^{i\Delta\beta_{12}z}A_1-\frac{\kappa_{23}}{\Delta\beta_{23}}e^{-i\Delta\beta_{23}z}A_3.
\label{eq3}
\end{equation}
From Eq.(\ref{eq3}), the AE condition namely $A_2$ being a dark sate is given by $
|\kappa_{12}|\ll|\Delta\beta_{12}|,\quad |\kappa_{23}|\ll|\Delta\beta_{23}|$. Using Eq.(\ref{eq1}) it simplifies to 
\begin{equation}
\left|\kappa\right|\ll \sqrt{(\beta^\prime-\beta_2^\prime)^{ 2}+\gamma^2}.
\label{eq4}
\end{equation}
We can conclude from Eq.(\ref{eq4}) that if the AE condition is satisfied for the Hermitian case \cite{xiangadi}, namely $\gamma=0$, it is automatically satisfied for the non-Hermitian case with $\gamma\neq0$. 

By making one waveguide a dark state, AE effectively reduces the number of the differential equations in (\ref{eqd})  from three to two. Replacing Eq.(\ref{eq4}) in Eq.(\ref{eqd}) and using Eq.(\ref{eq1}), one can write a Schr\"{o}dinger-like equation 
\begin{equation}
i\frac{d}{dz}\tilde{\psi}=-\mathcal{H}_{eff}\tilde{\psi},\quad\tilde{\psi}=(A_{+}, A_{-})^T
\label{eqtilde}
\end{equation}
where $A_{+,-}\equiv A_{1,3}\exp(\pm i\Delta \beta_{13}z/2)$. The elements of the Hamiltonian $\mathcal{H}_{eff}=\left(\begin{array}{cc}
\beta_{eff{+}}&\kappa_{eff}\\ \kappa_{eff}^\ast&\beta_{eff{-}}\end{array}\right)$ are 
\begin{equation}
\beta_{eff{\pm}}=\frac{|\kappa|^2}{\beta^\prime-\beta_2^\prime\pm i\gamma}\pm i\gamma,\quad \kappa_{eff}=\kappa_{13}+\frac{\kappa^2}{\Delta\beta_{12}}.
\label{eq6}
\end{equation}
From Eqs.(\ref{eq6}) and (\ref{eq4}), we observe that to the first-order approximation the imaginary part of the propagation constants are not affected by AB, while the couplings have been changed.

\begin{figure}
\includegraphics[width=1\linewidth, angle=0]{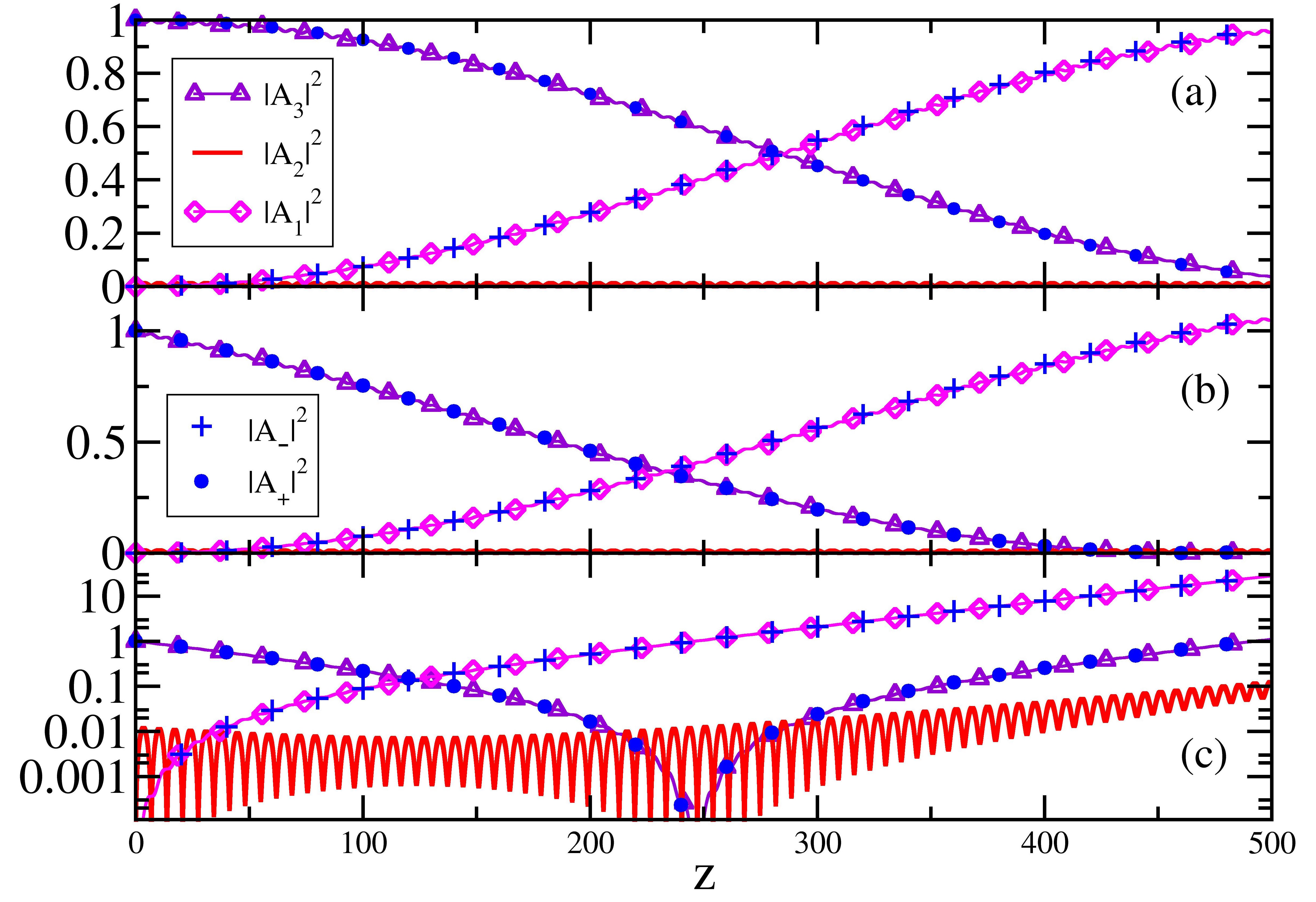}
\caption{(Color online) (a) Beam dynamics in a Hermitian trimer shown in Fig.(\ref{fig1}) in the AE regime with $\kappa_{13}=0$,$\kappa=0.05$, $\beta_2=1$ and $\beta_1=0.1$. The field from waveguide $A_3$ transfers into waveguide $A_1$ while intensity of the middle waveguide is close to zero. The curves indicated with $\bullet$ and $+$ are the results of integration of effective differential Eq. (\ref{eqtilde}). (b) Similar to (a) with non-zero $\gamma$ and in the exact phase. (c) Similar to (a) with $\gamma>\gamma_{PT}$. In this case AE is not preserved and the field intensity in the middle waveguide increases.\label{fig3}}
\end{figure}

We can use the $\mathcal{H}_{eff}$ to identify the exact phase of the coupled waveguides in the AE regime. Specifically, for $\kappa_{13}=0$ and $\kappa\in \Re$, the eigenvalues of $\mathcal{H}_{eff}$ are given by
\begin{equation}
{\mathcal E}_{1,2}=\frac{\kappa^2(\beta^\prime-\beta_2^\prime)\pm\sqrt{\Lambda}}{(\beta^\prime-\beta_2^\prime)^2+\gamma^2}.
\label{eq7}
\end{equation}
Where $\Lambda\equiv{(\beta^\prime-\beta_2^\prime)}^2 \kappa^4-\gamma ^2 \left({(\beta^\prime-\beta_2^\prime)}^2+\gamma ^2\right)^2+2 \gamma ^2 \kappa^2 \left({(\beta^\prime-\beta_2^\prime)}^2+\gamma ^2\right)$. Obviously, our structure is in the exact phase for $\Lambda>0$. In the upper panel of figure (\ref{fig2}), we plotted the real and imaginary part of the eigenvalues of $\mathcal{H}_{eff}$, namely $\mathcal E$, vs gain and loss parameter $\gamma$ for $\kappa=0.05$, $\beta^\prime=0.1$, $\beta_2^\prime=1$ and $\kappa_{13}=0$. The lower panel of Fig.(\ref{fig2}) depicts the spectrum, $\lambda$, of the Eq.(\ref{eqe} ) vs. $\gamma$. Three modes $\lambda_{1,2,3}$ are shown where two of them, $\lambda_{1,2}$, are in the main panel. The dark mode $\lambda_3$ is demonstrated in the sub-figure. A comparison between the upper and lower panels of Fig.(\ref{fig2}) shows that Eq.(\ref{eq7}) accurately predicts the spontaneous symmetry breaking point at $\gamma\equiv\gamma_{PT}=2.8\times 10^{-3}$. Thus, from numerical perspective, one can use the AE approximation to reduce the dimensionality of the system to identify the spontaneous symmetry breaking point.  Despite its accuracy in finding exceptional point, $\mathcal{H}_eff$ does not describe the spectrum of the coupled waveguides. We see a shift in the position of the two predicted modes $\lambda_{1,2}$ in the spectrum. In the next section, we show that this shift is not important and the dynamics is describes by the difference between the upper and lower levels, the so called bandwidth. Figure (\ref{fig2}) depicts that the bandwidth after AE approximation is preserved, $|\lambda_1-\lambda_2|=|\mathcal{E}_1-\mathcal{E}_2|=\frac{2\sqrt{\Lambda}}{(\beta^\prime-\beta_2^\prime)^2+\gamma^2}$. 

{\it Dynamics--} Armed with the previous knowledge about the eigenmode properties of the PT symmetric trimer in the AE regime, we move forward to study beam dynamics. The question is to determine whether the AE holds in each phase and how the dynamics is affected by the non-Hermiticity.

To answer the above question (see Fig.(\ref{fig3})) we excite the right waveguide of the system for different values of the gain and loss under the condition given in Eq.(\ref{eq4}) and check the electric field propagation along the $z$ direction. First, we consider the Hermitian case where $\gamma=0$. In this case, (\ref{fig3}a), the approximation that we used works with high accuracy and the Eq.(\ref{eqtilde}) shows the same dynamics as the actual system described by the set of differential equations in Eq.(\ref{eqd}). The excitation in the right waveguide in Fig.(\ref{fig1}) couples to the left waveguide while the intensity in the middle one remains almost zero. In figure (\ref{fig3}b), we introduce the gain and loss such that $0<\gamma<\gamma_{PT}$. We observe that while the overall dynamics is affected by the gain and loss as discussed in Ref.(\cite{hrbr}), the AE approximation describes the dynamics very well. Similar to the Hermitian case, the intensity in the middle waveguide remains zero. Finally in Fig.(\ref{fig3}c), we increase $\gamma>\gamma_{PT}$ such that the system enters into the broken phase. Although two sets of differential equations (\ref{eqd}) and (\ref{eq6}) predict the same dynamics, the intensity in the passive waveguide increases. Consequently, one can conclude that AE does not hold in the broken phase.

Physically, in the broken phase as depicted in the figure (\ref{fig2}), the imaginary part of one of the modes becomes positive and the other one becomes negative. Thus, in an exponential fashion one mode decays and the other one grows. In the spatial representation, the growing eigenmode has components in all three waveguides which results in an exponential growth in the total intensity of all waveguides. 

\begin{figure}
	\includegraphics[width=1\linewidth, angle=0]{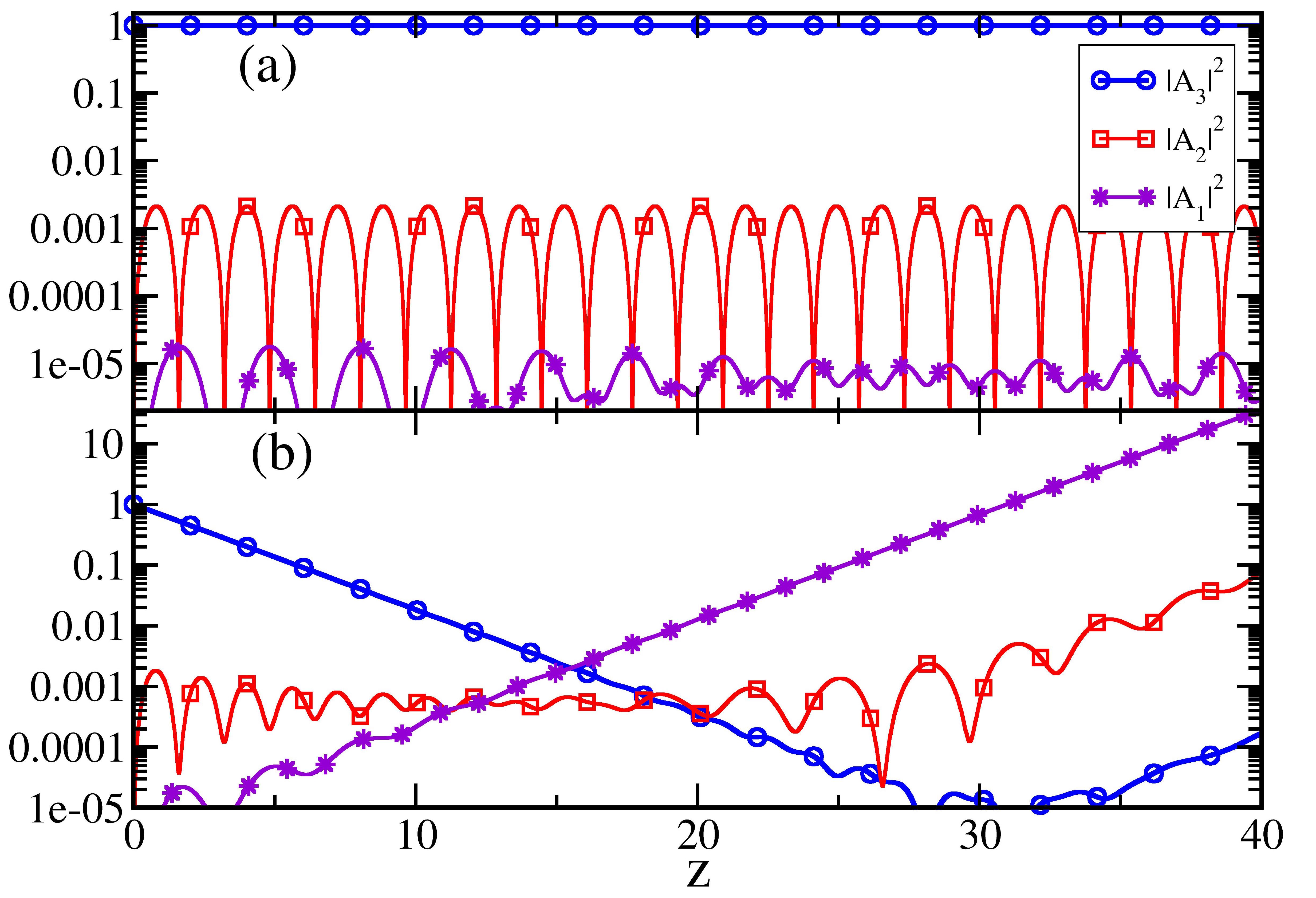}
	\caption{(Color online) (a) Beam dynamics in a Hermitian trimer shown in Fig.(\ref{fig1}) with broken P-symmetry. The parameters are chosen such that field intensity remains mainly in waveguide $A_3$. (b) Similar to (a) with non-zero $\gamma$. The system is in the broken phase and the complex eigenvalues of the system cause an exponential growth in the field intensity of the middle waveguide.\label{fig4}}
\end{figure}
While the above intuitive picture can be mathematically described, in the following we use a pedagogical and different approach to explain the breaking of AE in the broken phase.
Let us look for the condition in which the dark waveguide is either one of the outer waveguides. In this case from Equation (\ref{eqd}) and (\ref{eq2}), one finds that in addition to Eq.(\ref{eq4}) another condition must be fulfilled:
\begin{equation}
|V_{13}|\ll|\Delta\beta_{13}|
\label{eq9}
\end{equation}
If the system is Hermitian and P-symmetric, namely $\beta_1=\beta_3$, then the above condition cannot be satisfied. However, if the system does not preserve the P operator, in other words $\beta_{13}\neq 0$, then one can choose the parameters of the system such that equation (\ref{eq9}) is fulfilled. Figure (\ref{fig4}) represents the field intensities at each waveguide when we excited the waveguide with propagation constant $\beta_3$. Here, we chose the relative propagation constants to be $\beta_1=1$, $\beta_2=2$, $\beta_3=0.1$, and couplings to be $V=0.9$, $V_{13}=0$. These chosen parameters satisfy both conditions in Eq. (\ref{eq4}) and Eq.(\ref{eq9}), and thus, we see that the field remains in the excited waveguide with fluctuations of the order of $\sim10^{-3}$. We could make these fluctuation smaller by choosing larger values for $\beta_2$. However, this choice helps us to better understand the influence of the gain and loss on the dynamics of the field. 

As depicted in Fig.(\ref{fig4}), when we increase the gain and loss parameter $\gamma$ in the system, although we do not violate the condition (\ref{eq9}), the AE is no longer preserved and field intensity starts increasing in all waveguides. Note that because the real part of the propagation constants in the outer waveguides are not the same, the structure enters into the broken phase with any small but non-zero $\gamma$. The complex component of the eigenvalues amplifies the fluctuations and thus we observe that after 30 coupling units, the intensities in all waveguides starts increasing exponentially. Consequently, one can conclude that AE is not preserved in the broken phase.

{\it Conclusion--} We have shown that whilst AE is preserved in the exact phase, it does not hold in the broken phase where the intensity of the dark state exponentially increases. Our results become significant when we deal with coupled laser cavities or absorbers. Specifically, in such cases, using our proposed non-Hermitian AE effect, one can control the lasing output or absorbing efficiency by changing the external pump power or losses.

{\it Acknowledgments --} 
H.R gratefully acknowledge support from the UT system under the Valley STAR award.

%==================================================================================


\begin{thebibliography}{99}
\bibitem{1}	V. J. Sorger, Z. Ye, R. F. Oulton, Y. Wang, G. Bartal, X. Yin, and X. Zhang, Nature Communications 2, 331 (2011), doi:10.1038/ncomms1315

\bibitem{2}	S.M. Spillane, T.J. Kippenberg, and K.J. Vahala, Nature, 415, 621-623 (2002) doi:10.1038/415621a

\bibitem{disk}H. Ramezani, T. Kottos, V. Shuvayev, L. Deych
Physical Review A 83 (5), 053839 (2011)

\bibitem{eli} E. Yablonovitch, JOSA B 10 (2), 283-295 (1993)

\bibitem{Zheludev} Nikolay I. Zheludev, Yuri S. Kivshar, Nature Materials 11, 917–924 (2012) doi:10.1038/nmat3431

\bibitem{ramy}	K. G. Makris, R. El-Ganainy, D. N. Christodoulides, and Z. H. Musslimani, Phys. Rev. Lett. 100, 103904 (2008).
\bibitem{nat10}	C. E. Ruter, K. G. Makris, R. El-Ganainy, D. N. Christodoulides, M. Segev, and D. Kip, Nat Phys 6, 192 (2010).
\bibitem{flat} Hamidreza Ramezani, arXiv preprint arXiv:1701.00291 (2017)
\bibitem{elec}	J. Schindler, Z. Lin, J. M. Lee, H. Ramezani, F. M. Ellis, and T. Kottos, J. Phys. Math. Theor. 45, 444029 (2012).

\bibitem{prx} X. Zhu, H. Ramezani, C. Shi, J. Zhu, X. Zhang, Physical Review X 4 (3), 031042 (2014)

\bibitem{natcom} C. Shi, M. Dubois, Y. Chen, L. Cheng, H. Ramezani, Y. Wang, X. Zhang Nature communications 7 (2016)

\bibitem{newj} H. Ramezani, M. Dubois, Y. Wang, Y. R. Shen, X. Zhang
New Journal of Physics 18 (9), 095001 (2016)

\bibitem{alu} Romain Fleury, Dimitrios Sounas and Andrea Alu, Nature Communications 6 (2015)

\bibitem{laser1} H. Ramezani, S. Kalish, I. Vitebskiy, T. Kottos, Phys. Rev. Lett. 112 (4), 043904 (2014)

\bibitem{laser2}
H. Ramezani, H-K. Li, Y. Wang, X. Zhang,Phys. Rev. Lett. 113 (26), 263905 (2014)
\bibitem{absorber} H. Ramezani, Y. Wang, X. Zhang, IEEE Journal of Selected Topics in Quantum Electronics 22 (5) (2016)

\bibitem{bender} Carl M. Bender and Stefan Boettcher, Phys. Rev. Lett. 80, 5243 (1998)

\bibitem{zin} Z. Lin, H. Ramezani, T. Eichelkraut, T. Kottos, H. Cao, D. N. Christodoulides, Phys. Rev. Lett. 106 (21), 213901 (2011)

\bibitem{dim} A. Regensburger, C. Bersch, M-A.Miri, G. Onishchukov, D. N. Christodoulides, and U. Peschel, Nature 488, 167–171 (2012) doi:10.1038/nature11298

\bibitem{henning} H. Schomerus, N.Y. Halpern,Phys.Rev. Lett. 110 (1), 013903 (2013)
\bibitem{yang} B. Peng, S. K. Ozdemir, S. Rotter, H. Yilmaz, M. Liertzer, F. Monifi, C. M.  Bender, F. Nori, L. Yang, Science 346 (6207), 328-332 (2014)

\bibitem{hen} S. Malzard, C. Poli, H. Schomerus, Phys. Rev. Lett. 115 (20), 200402 (2015)
\bibitem{alex} S. Weimann, M. Kremer, Y. Plotnik, Y. Lumer, S. Nolte, K. G. Makris, M. Segev, M. C. Rechtsman, and A. Szameit, Nat. Mater. 16, 433–438 (2017) doi:10.1038/nmat4811

\bibitem{xiangadi} M. Mrejen, H. Suchowski, T. Hatakeyama, C. Wu, L. Feng, K. O Brien, Y. Wang, and X. Zhang, Nature Communications 6, 7565 (2015)
doi:10.1038/ncomms8565
\bibitem{salamo} A. Guo, G. J. Salamo, D. Duchesne, R. Morandotti, M. Volatier-Ravat, V. Aimez, G. A. Siviloglou, and D. N. Christodoulides, Phys. Rev. Lett. 103, 093902 (2009)
\bibitem{hrbr} H. Ramezani, J. Schindler, F. M. Ellis, U. Gunther, T. Kottos
Physical Review A 85 (6), 062122 (2012)

\end{thebibliography}
\end{document}